\documentclass[sigconf]{acmart}

\AtBeginDocument{%
  \providecommand\BibTeX{{%
    \normalfont B\kern-0.5em{\scshape i\kern-0.25em b}\kern-0.8em\TeX}}}

\copyrightyear{2025}
\acmYear{2025}
\setcopyright{cc}
\setcctype{by}
\acmConference[CHI '25]{CHI Conference on Human Factors in Computing Systems}{April 26-May 1, 2025}{Yokohama, Japan}
\acmBooktitle{CHI Conference on Human Factors in Computing Systems (CHI '25), April 26-May 1, 2025, Yokohama, Japan}\acmDOI{10.1145/3706598.3713153}
\acmISBN{979-8-4007-1394-1/25/04}

\acmSubmissionID{7845}

\usepackage{fontawesome}
\usepackage{color}
\usepackage{colortbl}
\usepackage{xcolor}
\usepackage{makecell}
\usepackage{multirow}
\usepackage{multicol}
\usepackage{xspace}
\usepackage{hyperref}
\usepackage{balance}

\usepackage{graphicx}
\usepackage{etoolbox}
\usepackage{algorithm}
\usepackage{algpseudocode}
\usepackage{amsmath}
\usepackage{amsfonts} 
\newcommand{\name}{\textsc{Typoist}\xspace}
\newcommand{\benchmark}{\textsc{TypingError}\xspace}

\definecolor{bad}{rgb}{0.99608,0.87843,0.82353}
\definecolor{good}{rgb}{0.89804, 0.96078, 0.87843}
\definecolor{best}{rgb}{0.63137, 0.85098, 0.60784}

\usepackage{ifthen}
\newboolean{revising}
\setboolean{revising}{false}
\ifthenelse{\boolean{revising}}
{
    \newcommand{\rv}[1]{\textcolor{blue}{#1}}
} {
    \newcommand{\rv}[1]{    #1}
}

\begin{document}

\title{Simulating Errors in Touchscreen Typing}


\author{Danqing Shi}
\orcid{0000-0002-8105-0944}
\affiliation{\institution{Aalto University}
\city{Helsinki}
\country{Finland}}

\author{Yujun Zhu}
\orcid{0000-0001-7119-6328}
\affiliation{\institution{Aalto university}
\city{Helsinki}
\country{Finland}}

\author{Francisco Erivaldo \\Fernandes Junior}
\orcid{0000-0003-2301-8820}
\affiliation{\institution{Aalto University}
\city{Helsinki}
\country{Finland}}

\author{Shumin Zhai}
\orcid{0000-0003-0752-2090}
\affiliation{\institution{Google}
\city{Mountain View}
\country{USA}}

\author{Antti Oulasvirta}
\orcid{0000-0002-2498-7837}
\affiliation{\institution{Aalto University}
\city{Helsinki}
\country{Finland}}

\renewcommand{\shortauthors}{Shi, et al.}

\begin{abstract}
Empirical evidence shows that typing on touchscreen devices is prone to errors and that correcting them poses a major detriment to users’ performance. Design of text entry systems that better serve users, across their broad capability range, necessitates understanding the cognitive mechanisms that underpin these errors. However, prior models of typing cover only motor slips. The paper reports on extending the scope of computational modeling of typing to cover the cognitive mechanisms behind the three main types of error: slips (inaccurate execution), lapses (forgetting), and mistakes (incorrect knowledge). Given a phrase, a keyboard, and user parameters, ~\name simulates eye and finger movements while making human-like insertion, omission, substitution, and transposition errors. Its main technical contribution is the formulation of a supervisory control problem wherein the controller allocates cognitive resources to detect and fix errors generated by the various mechanisms. The model generates predictions of typing performance that can inform design, for better text entry systems.
\enlargethispage{20pt}\end{abstract} 

\ccsdesc[500]{Human-centered computing~HCI theory, concepts and models}

\keywords{Human errors; User simulation; Mobile typing}

\maketitle

\section{Introduction}

\begin{figure*}[!t]
\centering
  \includegraphics[width=\textwidth]{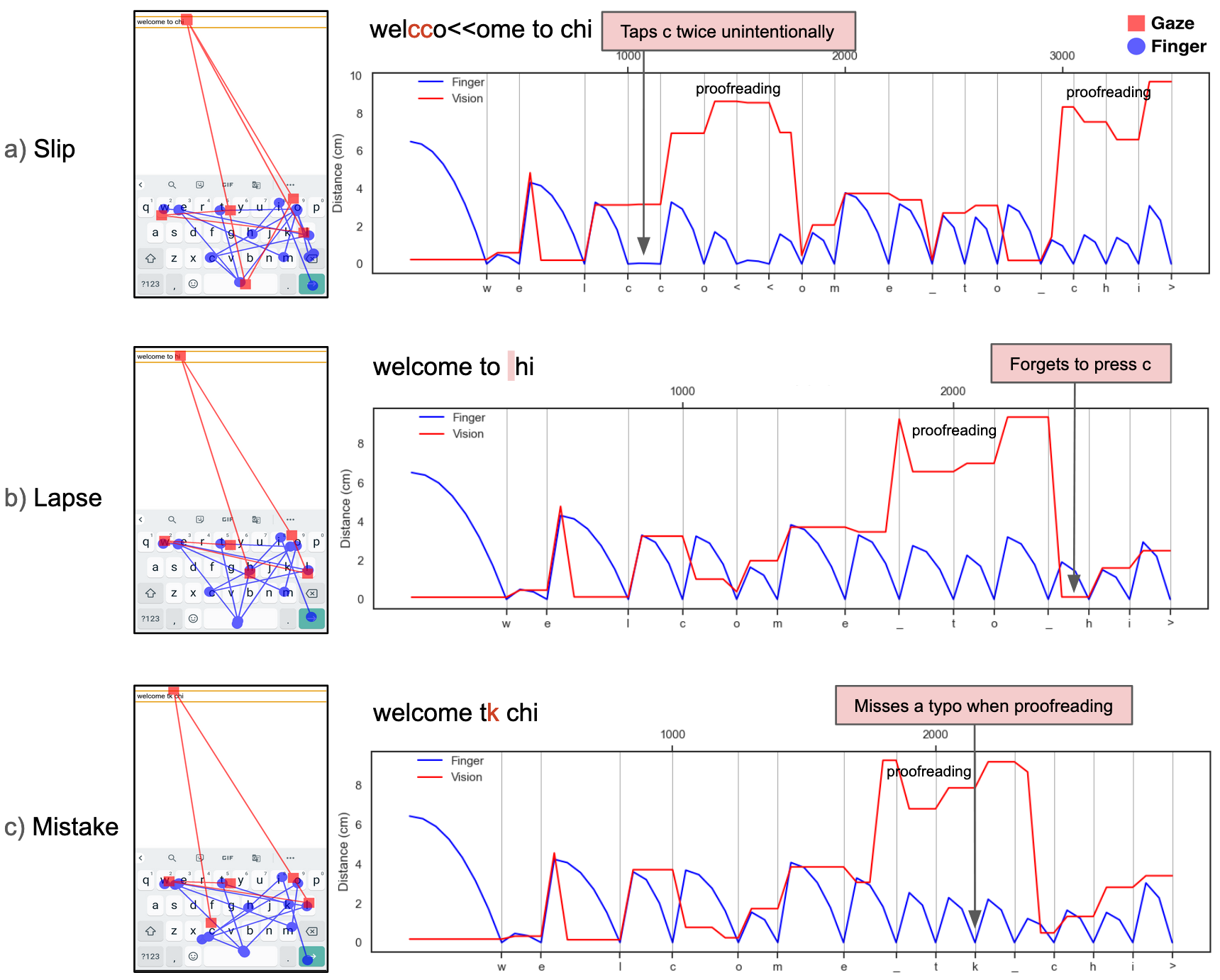}
  \caption{
    We introduce the first model covering a wide spectrum of errors known to be commonplace in typing, ~\name, which simulates the way users move their eyes and fingers when they type. The figure illustrates the three main types of error covered by the model. a) Slip: accidentally double-tapping while typing rapidly, which is detected through proofreading and corrected by backspacing; b) Lapse: forgetting where the finger was. c) Mistake: missing a typo and believing it is correct.}
  \label{fig:teaser}
\end{figure*}

\rv{
Human beings make errors in all lines of work and spheres of life~\cite{donaldson2000err}. 
In touchscreen typing, human error creates a major hindrance to performance, one that manifests itself very differently between users~\cite{dhakal2018observations}.
Typing performance depends greatly on how quickly users can type and confirm the typed text, a process that errors can disrupt. Among common errors are hitting the wrong key, repeating a letter, forgetting what has been typed, and overlooking grammar mistakes.
The ``fat finger problem''~\cite{siek2005fat} is a well-known issue in typing. The term refers to the increased likelihood of hitting the wrong key when the keys are small.  As a result, users often have to slow down their movements to touch the screen more precisely~\cite{bi2013ffitts}.
Advanced features like autocorrect can assist with error management, but they may also significantly alter user behavior~\cite{banovic2019limits}. 
Detecting errors, deciding whether to correct them, and implementing corrections involve complex interactions of perceptual, cognitive, and motor control processes~\cite{jiang2020we}.
}
Not surprisingly, errors are perhaps the single most critical factor constraining typing performance ~\cite{palin2019people}. 
To improve text entry systems, we need to understand what causes errors. 

\emph{Computational models} have advanced both theorizing and practical efforts in the text entry domain.
They have explained how limitations of the human motor system lead to inaccuracies in input~\cite{zhai2004characterizing} and address how users exercise some strategic control over such errors \cite{guiard2015mathematical}.
Models of typing are of practical value too.
They have been used to optimize layouts, drive intelligent text entry techniques,  and personalize keyboards (e.g., \cite{bi2013bayesian,weir2014uncertain,zhai2000metropolis}).
However, even the most comprehensive simulation-based models have been limited mainly to a single error type: motor slips \cite{jokinen2017modelling,jokinen2021touchscreen, shi2024crtypist}.
This restricts these models' usefulness -- they need to be expanded.
In particular, they should cover the key mechanisms behind errors and their typographical consequences. 
The prevailing conception is that errors in typing, as do human errors in general, fall into three main types~\cite{reason1990human}: \textit{slips}, which occur when motor execution deviates from the intended outcome; \textit{lapses}, which are due to memory failures; and \textit{mistakes}, arising from incorrect or partial knowledge (see Figure~\ref{fig:teaser}).

The goal of this paper is to shed new light on mechanisms possibly underlying typing errors and, thereby, significantly increase the scope and realism of models' predictions. 
Our key insight is that the mechanisms that produce errors are partially under strategic control. 
Users almost always have \emph{some} strategic control over the probability of these.
They can reduce errors by allocating more time and resources. 
They can slow down and monitor what they do more closely.
Yet they might not always want to do that.
What is optimal for them depends on their preferences.
A user who just wants to send a message quickly may not care about possible typos. 
We want to capture this critical supervisory aspect of errors.
\rv{
One recent model~\cite{jokinen2021touchscreen} has sparked efforts to explore eye--hand coordination in touchscreen typing, illuminating how users decide to assign visual attention. 
These culminated in the latest model, CRTypist~\cite{shi2024crtypist}, a supervisory control model that reproduces people's cognitive processes, for more human-like touchscreen typing behavior.
Yet these previous models still only cover motor slips.
}

In this paper, we present ~\name, which contributes to capturing the interplay between two types of cognitive mechanism: those that produce errors and those that attempt to detect and fix them.
First, we cover three cognitive mechanisms that produce errors in typing.
~\name adds two sources of error: memory and perception. 
Our model's motor control system may forget commands sent to it;
likewise, its representation of vision is noisy and not always able to detect errors in text even upon reading it.
Secondly, 
the model exploits broader-based handling of the supervisory control process.
We model how it decides to allocate resources to two critical subtasks of typing: moving the fingers and checking the text (proofreading).
In contrast, information-processing-based models of cognition put less emphasis on the closed-loop aspect of errors ~\cite{wickens2021engineering}.
Specifically, we assume that supervisory control proceeds from \emph{beliefs} formed from perceptual samples.
At any given time, the model is looking at some portion of the display, thus obtaining a sample.
The beliefs formed depend on whether the agent has looked at the text that has been typed, its fingers, or the keys. Each such belief itself is, in turn, constrained: subject to forgetting.
Putting these two fundamental mechanisms together, our model can cover four main types of typographical errors: insertion, omission, substitution, and transposition errors.

\name exhibits practical utility. 
It can be given a phrase, a keyboard design, and assumptions about the user. From these, it simulates how the given user would type text while moving the eyes and fingers, and creating errors. 
At the moment, it covers standard touchscreen keyboards that utilize the QWERTY layout and offer an autocorrection feature.
We found that users with different capabilities can be simulated by changing the values of parameters that describe their capabilities. 
We offer an interactive tool for testing out parameter values and an optimization method for inferring parameters from a dataset.

To test the validity of the model, 
we developed the \benchmark benchmark, which involves diverse user groups, including young adults, users with Parkinson's, elderly users, and individuals using different types of keyboards, all of whom vary in the error types and frequencies faced. 
The model proved able to replicate key characteristics of real  error distributions, whereas the prior state-of-the-art model reproduces merely a small subset of the errors.
Furthermore, \name was able to capture the distribution of error correction across typing speeds, thus demonstrating an ability to account for individual-to-individual differences. When judged in light of in-lab gaze and finger movement data, it exhibited more accurate, realistic behavior in error handling and correction strategies than the baseline model.
Finally, it generated human-like behavior in scenarios that incorporate using autocorrection. 

In summary, the main contribution of our research is the extension of computational modeling of typing to simulate the majority of common typing errors in a touchscreen typing environment.  
\rv{
The advantages \name possesses over the latest typing model, CRTypist~\cite{shi2024crtypist}, include the following:
1) integrating three error-generating mechanisms (slips, lapses, and mistakes) into a unified model;   
2) upgrading the supervisory control model to support detecting and correcting diverse errors; and
3) improving the computational rationality modeling workflow, through joint parameter optimization, to achieve more human-like performance.
However, \name does not cover real-world behaviors involving advanced features. More work is needed to better handle complex dynamics in touchscreen typing.
\name is available via \texttt{\url{https://typoist.github.io/}}.
}

%


\section{Related Work}

For background, we begin by reviewing related work on how humans make and correct typing errors. We then examine human-error modeling approaches and highlight the research gap.

\subsection{Typographical errors}

\emph{Typographical errors} are errors in typed or printed text. 
Several definitions have been developed to suit text entry contexts,
most of which originate from studies of typing with physical keyboards. 
\rv{
Three fundamental sorts of character-level error take place in typing~\cite{wobbrock2007measures}:
\emph{Insertion errors} occur when an extra character is typed by mistake (e.g., \texttt{typist} becoming \texttt{typ\textbf{o}ist}).
\emph{Omission errors} arise when a character is missing from the word typed (e.g., \texttt{typist} turning into \texttt{tpist}).
\emph{Substitution errors}, also known as \emph{misstrokes}, account for a large proportion of the errors people make when typing a nearby character (e.g., \texttt{typist} ends up as \texttt{typ\textbf{u}st}).
All these types can be identified well through character-level error analysis based on minimum string distance~\cite{mackenzie2002text}. 
Although they together characterize a large percentage of typing errors~\cite{gentner1983glossary}, other errors too appear frequently in typing.
For instance, a \emph{transposition error}~\cite{rumelhart1982simulating} arises from reversing two adjacent characters while typing (e.g., \texttt{should} becomes \texttt{shou\textbf{dl}}); 
a \emph{doubling error} occurs when a word that contains double letters gets the wrong letter doubled (e.g., \texttt{look} turns into \texttt{lo\textbf{kk}}); and there are 
\emph{alternation errors}, similar to doubling errors but with an alternative sequence of characters (e.g., \texttt{these} becomes \texttt{th\textbf{ses}}). 
This paper follows the categorization proposed by~\citet{wang2021facilitating} for computing error metrics, including the error rates of insertion, substitution, omission, and transposition.
}

\subsection{Errors in touchscreen typing}

Typing on touchscreens is known to be error-prone ~\cite{hoggan2008investigating,palin2019people}. 
Accordingly, large datasets have been examined for their distributions of typographical errors \cite{palin2019people}.
Some of the errors are attributed to the difficulty of hitting small keys ~\cite{holz2010generalized, holz2011understanding}, 
connected with the aforementioned fat finger problem~\cite{siek2005fat,baudisch2009back}. 
There are also large differences arising from personal characteristics. 
For instance, one study found that omission, primarily with cognitive causes, 
is the most common error type among elderly individuals~\cite{nicolau2012elderly}. 
Another study revealed that users with Parkinson's make significantly more insertion, substitution, and omission errors~\cite{wang2021facilitating}. 
The researchers' analysis attributed this to hand tremors creating a longer distance between adjacent touches~\cite{holz2011understanding}.
Also, people's error rates change situation-specifically. More typos occur when they feel free to leave errors~\cite{palin2019people} as compared to when they are trying to avoid errors~\cite{jiang2020we}.
Our research aim was to replicate the tendencies reported in the general population as well as those of two specific user groups.

Importantly for our work, typographical approaches to errors manifest a crucial limitation.
Most prominently, modern keyboards introduce interactive features and, thereby, novel types of errors that cannot be understood as typographical ones.
For example, users making \emph{mode errors} have hit the correct keys but in an inappropriate mode, such as with Caps Lock on ~\cite{norman1988psychology} (e.g., \texttt{hello} becomes \texttt{HELLO}).
Another kind, \emph{autocorrection errors}, occurs especially frequently with mobile devices when a statistical decoder incorrectly assumes the desired word to be one different from that intended. 
When a relatively effective autocorrection feature is in place and the user is less concerned about errors, people engage in faster typing and movement between keys~\cite{banovic2017quantifying, banovic2013effect}.
Therefore, we strove to encompass autocorrection too.

\subsection{How users detect and fix errors}

The ability to detect errors is essential for high performance in typing. 
There are two ways a user can detect that an error has been made:
1) by recognizing it in the text while reading and
2) by noticing an erroneous keypress~\cite{logan2010cognitive}. 
When it comes to the former, visual attention has an important role.
Of course, users cannot look at their fingers the whole time, since they need to check the text display also.
Researchers found that typists keep their gaze on the keys about 70\% of the time when typing with one finger and about 60\% of the time while using their thumbs ~\cite{jiang2020we}.
They glance at the typed text about four times per sentence (avg. sentence length: 20 characters).
There is a further layer here: errors can occur even with proofreading.
If nothing else, some incorrect text may not be properly detected~\cite{haber1981error}. 
The speed with which users are able to perform proofreading is associated with the accuracy of error detection~\cite{pilotti2009text}.
We have modeled the associated accuracy to cover this.

After detecting an error, the user may choose to fix it.
Evidence suggests that there are two strategies to this end ~\cite{pinet2022correction}: 1) immediate backspacing to correct the error and 2) delayed corrections~\cite{arif2009analysis, jiang2020we}.
Furthermore, users can influence the probability of errors strategically.
They might slow down to hit keys more precisely \cite{bi2013ffitts},
one might develop a strategy of moving (two) fingers to minimize the fat finger problem and reduce rapid repetitive motions of any one finger~\cite{cerni2016}, etc.
We aimed to model such strategies. 

\subsection{Models of mechanisms behind typing errors}

Most prior work has focused on errors caused by the limitations of the human motor control system.
It has shown that Fitts' law functions well for predicting user performance in pressing keys under various conditions \cite{zhai2004characterizing}.
This model, often employed for examining slips during typing ~\cite{wobbrock2008error},
was extended with Finger Fitts' law~\cite{bi2013ffitts} to consider predictions for touchscreen typing. 
Modeling of this nature is limited to motor slips; however,
recent work has started to look at simulation-based approaches that could cover eye--hand coordination during touchscreen typing~\cite{jokinen2021touchscreen}. 
This paves the way toward predicting detail-level gaze and finger movement behavior in proofreading and error correction.

\rv{
CRTypist~\cite{shi2024crtypist} represents the latest model in this area. Its design is based on a supervisory control framework that includes three key modules: vision, finger, and working memory, each trained to replicate human cognitive processes. 
The vision module manages visual attention, enabling the model to shift focus between the keyboard and the text display for proofreading and finger guidance. The finger module simulates motor control, allowing for tapping keys on a touchscreen. Finally, the one for working memory maintains a time-decaying belief about the typed text, which informs decision-making for subsequent actions. Overarching supervisory control coordinates these modules to optimize typing performance, balancing between speed and accuracy. 
This modular, hierarchical design enables effectively predicting typing performance across various designs, tasks, and user groups, but the model focuses mainly on finger movement accuracy, so it addresses just substitution errors related to motor slips. A significant gap remains: simulating other types of errors that can occur during typing.
}

\begin{figure}[!t]
\centering
  \includegraphics[width=0.49\textwidth]{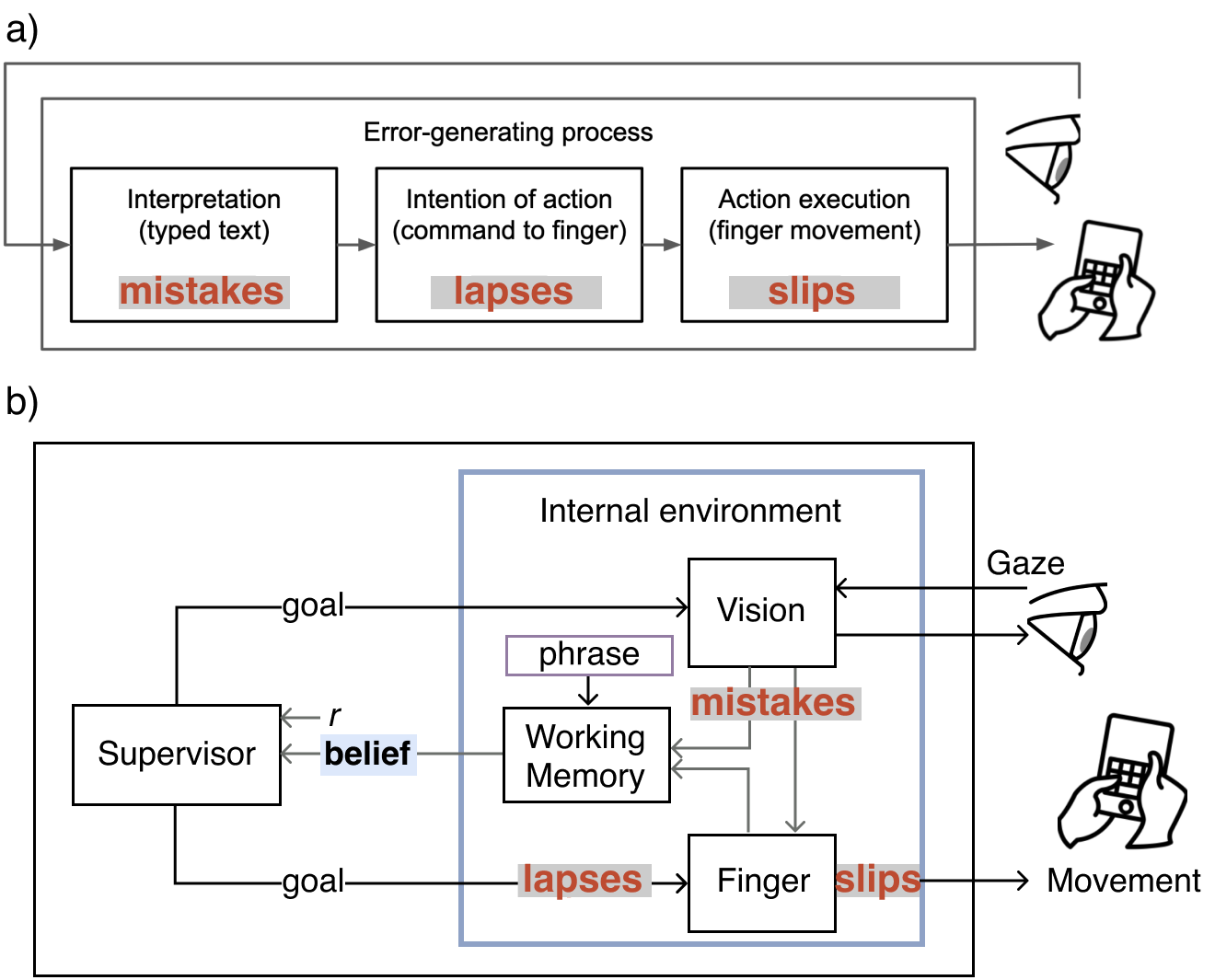}
  \caption{
  a) An information-processing view of human error ~\cite{wickens2021engineering} assumes that a typing error can be produced by any step in a sequence of three: interpretation, intention, and execution. 
  Slips are incorrectly executed movements,
  lapses are incorrect commands, and mistakes emerge when misinterpretation of the typed text leads to inappropriate decisions about what to do.
  b) ~\name extends the architecture that underpins \textsc{CRTypist}. With ~\name, the system models the cognitive processes that generate errors. Moreover, the supervisory controller can observe the consequences of errors~\cite{shi2024crtypist}.}
  \label{fig:model}
\end{figure}

\section{\name: The Modeling Principles and Design}

The primary goal for ~\name is to reproduce human-style typing errors, including the way people make corrections \cite{pinet2022correction}, without compromising the overall realism of the model's predictions relative to the previous state-of-the-art model~\cite{shi2024crtypist}. In addition, we wanted the model to be able to run directly on pixels as in previous work 
and to account for individual differences. To this end, our modeling approach employs three principles, discussed below.
The first marks the most significant advance, and the other two are assumptions that, while developed in prior work,
we have adapted to account for more types of errors. All three are drawn together into a single model.

\paragraph{Noisy cognitive capabilities.} 
We model cognitive resources as limited-capacity channels. 
When these resources are requested to be faster, they generate more errors.
While previous work has applied this principle to model motor control in typing, we here extend it to cover vision and working memory.
Specifically, \textsc{Vision} controls the gaze movement to observe the screen, processing pixels through foveated and peripheral views; \textsc{Working memory} holds information about what has been typed with a level of uncertainty.
An important feature of our model of these capabilities is that they contain theory-inspired empirical parameters that contain the level of noise. 
This allows us to simulate users with different abilities.

\paragraph{Hierarchical supervisory control.} 
We assume that there are two levels of control in typing: high and low.
Higher-level control takes a supervisory role~\cite{botvinick2012hierarchical, frank2012mechanisms}. 
It monitors what happens (based on its beliefs) and sets goals accordingly for low-level controllers.
At the low level, two motor systems are responsible for movement: one handles eye control, and the other controls the fingers.
These systems are given a goal (e.g., to press ``K''), which they try to reach in a way that factors in their own, limited abilities.
This hierarchical approach confers greater modeling power: we can now model these abilities independently of each other, as opposed to in an end-to-end manner.
It also gives a boost to training, because we can train the controllers separately.

\paragraph{Computational rationality.} 
The final assumption is that the (high- and low-level) controllers adjust their policy to maximize expected utility,
while optimality is bounded by the noisy cognitive abilities \cite{oulasvirta2022computational}. 
In practice, that entailed formulating typing as a partially observable Markov decision process (POMDP). This is consistent with prior work \cite{jokinen2017modelling,jokinen2021touchscreen},
but we added a new element by introducing error-producing mechanisms in the cognitive environment of the supervisory controller. 

\subsection{Noisy cognitive capabilities}
\label{sec:errors-generating}

\rv{
One key contribution of the model lies in covering diverse noisy cognitive capabilities in a unified model, whereas previous modeling studies considered only a subset of finger slips.
}
Slips, lapses, and mistakes are connected with different but partially overlapping generation mechanisms~\cite{reason1990human}. 
Slips are unintended and uncontrolled actions, lapses occur when people forget to do something, and mistakes are incorrect decisions that a person makes in the mistaken belief that this is the right thing to do. 

We took an information processing approach to categorizing human errors~\cite{wickens2021engineering}, then applied the resulting framework to touchscreen typing. 
Our mapping from the information processing perspective to transcription typing is illustrated in Figure~\ref{fig:model}~(a), where each component leads to one of the specific types of human error, which are interconnected into a complete process. 
When interacting with a touchscreen, individuals may gain inaccurate perceptions of the text typed, which lead to mistakes in their typing. Subsequently, they might forget to perform corrective typing actions, because of memory lapses. Finally, slips in motor control can cause them to execute finger movements incorrectly. 

Rather than list every possible error in each category, we adhered to Occam's razor and identified the major factors in the human errors that occur often in touchscreen typing. In our model, each latent mechanism at play in these errors is controlled by at least one error parameter, for factoring in the relevant capability. Through combining these mechanisms, the model can replicate diverse human errors.

\subsubsection{Slips}

Slips happen when there is a discrepancy between intention and execution. In touchscreen typing, slips are often caused by motor control errors due to physical limitations such as hand tremors or the fat finger problem.
The precision of fingertip movement depends on motor control noise, which varies with speed and distance ~\cite{fitts1954information}.

We simulate this underlying mechanism by using the Weighted Homographic (WHo) model~\cite{guiard2015mathematical}: $(y-y_0)^{1-k_\alpha}(x-x_0)^{k_\alpha} = F_K$,
In this model, $x$ represents the movement time of the finger, $y$ represents the standard deviation for the spread of the finger's endpoint, and $F_K$ is a parameter that controls finger capability – a smaller $F_K$ value indicates more accurate movement. This motor control noise can lead to substitution errors (tapping a key adjacent to the intended one etc.) or omission errors (the finger not hitting any key). 

We simulate other types of slips also -- specifically, unintentional double taps and swapping of motor commands, which are influenced by finger movement speed: $P(v)=1 - e^{-k \cdot v}$. Higher typing speed can increase the likelihood of unintended insertions and transpositions.
In the case of double tapping, the finger makes a movement to the same key immediately, while swapping of motor commands can disturb the keystroke order when the finger is close to the key that should come \emph(after) the next one.

\subsubsection{Lapses}

In touchscreen typing, lapses occur when people forget to give a command to their fingers, such that steps in the process get skipped. These mistakes are often attributed to cognitive errors resulting from forgetfulness~\cite{nicolau2012elderly}.

\name simulates this latent mechanism by modeling the probability of forgetting to give a motor command to the fingers at character level. That is, we assume that, when people's memory of what has been typed is weak, they could forget to type what they intended to type next.
We simplify the likelihood of this by randomly forgetting a character to type, related to the time $t$ since the last proofreading, using exponential decay: $P(t) = 1 - e^{-kt}$, where $k$ is a free parameter that controls the likelihood of forgetting. With a lower $k$ value, fewer lapses occur during typing, with a minimum of $k = 0$, at which there are no lapses.

\subsubsection{Mistakes}

In touchscreen typing, mistakes can be attributed to incorrectly observing the touchscreen. This has two aspects: misreading already-typed text during proofreading and inaccurately observing the finger's position during visual guidance.
The first mechanism is related to the accuracy of proofreading. It is possible for a user focusing on the text field to overlook errors and perceive incorrect text as correct. This affects error handling. We model the mechanism by expressing the conditional probability of missing a typo during proofreading via the time-dependent function $P_{\text {obs--text}}= p_0 \cdot e^{-T}$, where longer-duration proofreading increases the likelihood of accuracy.

The second mechanism manifests itself during visual guidance when the gaze is on the finger. Occlusion may lead to inaccurate observation of the finger's position~\cite{baudisch2009back}; that is, a finger obstructing some part of the keyboard could make it difficult to determine the position accurately. We use a constant value $P_{\text {obs--finger}}$ to model the conditional probability of missing a finger slip caused by finger movement during visual guidance.

\begin{figure*}[!t]
\centering
  \includegraphics[width=\textwidth]{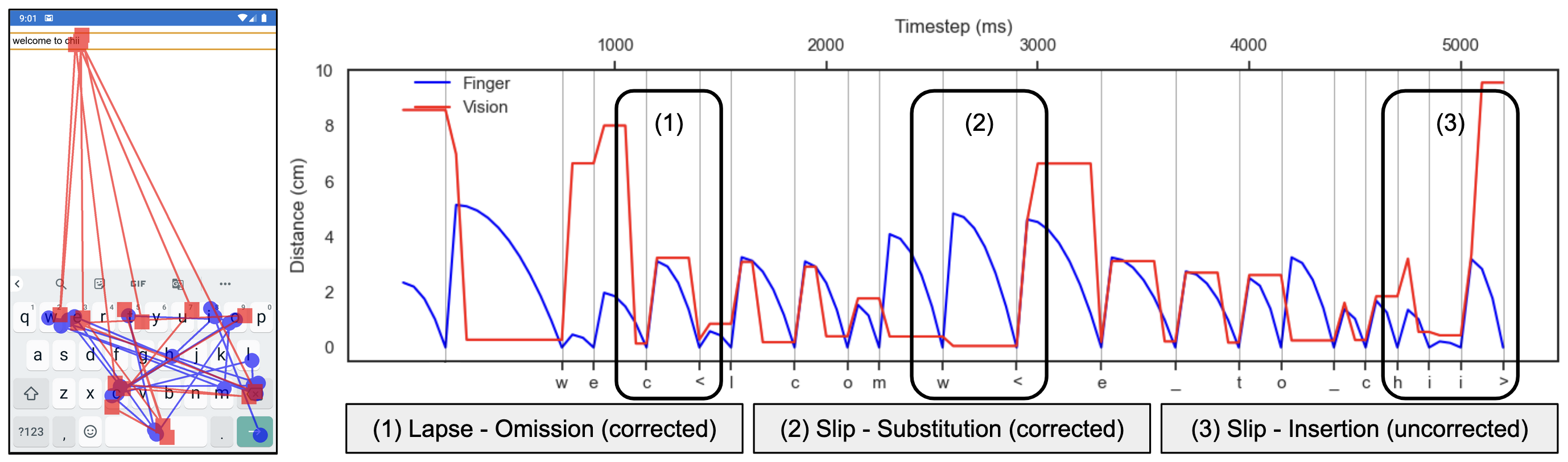}
  \caption{A simulation example involving multiple mechanisms that generate various text errors and corrections. In typing of ``welcome to chi'' with the Gboard interface, the following errors occur: 1) the model initially forgets to type the letter ``l'' (an omission error) though then quickly correcting it; 2) it accidentally types ``e'' instead of ``w'' (making a substitution error) although it corrects this mistake as well; 3) and, at the end of the sentence, it makes an insertion error by double tapping ``i'' -- with the model failing to detect this and submitting the text as-is.}
  \label{fig:errorcase}
\end{figure*}

\subsection{\rv{Hierarchical supervisory control}} 
\label{sec:supervisory-control}

People can strategically modulate the resources they allocate to precluding or correcting errors ~\cite{anderson2004integrated, fodor1983modularity}.
\rv{Our model's architecture design is anchored in that of the latest supervisory typing model, CRTypist~\cite{shi2024crtypist}, which models the supervisory control problem as deciding where to look and where to move the finger. Specifically, we built \name on the internal environment of CRTypist, which furnishes the interface between the control policy and the touchscreen. The three key components within this internal environment each have distinct abilities and limitations:}
\textsc{Vision} is responsible for moving the gaze to observe the screen from pixels via foveated and peripheral views; the \textsc{finger} decides on the finger movement for tapping on the touchscreen keyboard; and \textsc{working memory} holds both the information about what has been typed and the belief data. In general, the modeling for the first two of these is controlled by the supervisor in parallel in line with the belief from the memory.
\rv{
As is illustrated in Figure~\ref{fig:model}~(b), where our model diverges from the design of CRTypist is in integrating noisy cognitive capabilities into the supervisory control architecture by adding mistakes to the vision implementation for proofreading and visual guidance, adding lapses to the commands to the finger module, and adding slips to execution by the finger module – with all these error mechanisms being parameterizable factors that affect the internal environment. The parameters' effects are reflected further in the belief tied to the observation retrieved from working memory.
}

Putting it all together, we model typing with errors as a POMDP.
The supervisory controller attempts to type a phrase given to it.
However, it has only partial access to the touchscreen through the internal environment. 
Moreover, the internal environment is stochastic, arising directly from the error-creating mechanisms we describe above.
The POMDP definition is as follows:
\begin{itemize}
    \item The full state, $\mathcal{S}$, includes all information about the screen, at pixel level. This cannot be directly observed.
    \item The observation space, $\mathcal{O}$, supplies the belief as to what has been typed and the probabilities for each error type: the probability of missing a typo when proofreading, the probability of missing a finger slip, the probability of forgetting a motor command, that of an unintentional double tap, the probability of unintentional swapping of motor commands, and finger motor control noise. The reason we include these error-related beliefs in our observations is to make sure the model is able to adjust its behavior to the error capacity.
    \item The action space, $\mathcal{A}$, dictates the goals for both finger and gaze movements. Specifically, the goal for the finger is to reach the next key to be typed, while the vision's focus is split between the key and the input field. Once the goals are set, the vision and finger modules within the internal environment execute the actual movements, using pre-trained models introduced in previous work ~\cite{shi2024crtypist}.
    \item The reward function, $\mathcal{R} = (1 - \textit{Err}^{\alpha}) - w \cdot t$, combines error rates and the time budget, where $\textit{Err}$ represents the error rate, $\alpha$ controls sensitivity to errors, $w$ is the weight assigned to time, and $t$ is the time taken. This formulation encourages a balance between speed and accuracy.
\end{itemize}

\subsection{\rv{Computational rationality}}

\label{sec:optimization}
We followed the main steps of workflows geared for building computationally rational models of human behavior~\cite{chandramouli2024workflow}, where the goal is to train an agent to replicate human decision-making processes as closely as possible. In our case, the agent’s optimal policy for the supervisory controller is trained via reinforcement learning (RL) with Proximal Policy Optimization (PPO) from the \texttt{stable-baselines3} library~\cite{schulman2017proximal}, over the course of 5 million timesteps. During this training, the agent learns to predict human typing patterns by continually refining its policy in response to observed behaviors in the simulated environment. We chose PPO for its ability to effectively balance exploration and exploitation during training while also guaranteeing stability through its clipped objective function, which limits large, destabilizing policy updates \cite{schulman_proximal_2017}.

\rv{
Another improvement introduced by \name is parameter fitting for a computationally rational model through joint optimization. 
The goal is to achieve an optimal and stable policy within a large behavior space that accommodates diverse error-relevant behaviors.
Beyond the optimization of human parameters, the behavior of the model depends on the hyperparameters of the model's training. 
Careful selection of such parameters is essential, since they significantly influence the performance of RL agents ~\cite{andrychowicz_what_2020, paine_hyperparameter_2020, yang_efficient_2021}, and even small changes in the implementation of RL algorithms can affect their performance~\cite{engstrom_implementation_2020}.
}

To infer the optimal parameters, we used a two-loop optimization process to jointly optimize parameters. In the outer loop, the model is trained with a variety of human parameters, while in the inner loop, it identifies the optimal user group characteristics that the agent can model. 
This process seeks to pinpoint the best combination of model hyperparameters and human parameters, in order to optimize the typing model.

\begin{itemize}
    \item \textit{Outer loop optimization}: The outer loop focuses on optimizing key hyperparameters that influence the process for training the RL agent (e.g., the entropy coefficient and clipping range). Optimizing these hyperparameters is important because they directly affect how the agent interacts with its environment and learns from that interaction. In the outer loop, these hyperparameters are refined to minimize the difference between the agent's typing behavior and the target human typing behavior, measured in terms of the Jensen--Shannon divergence~\cite{shi2024crtypist}. This loop aims to find a general typing model that works well across the full range of user behaviors.
    \item \textit{Inner loop optimization}:  Within each iteration of the outer loop, the inner one optimizes the parameters that are essential for adapting the agent to distinct user groups. These parameters reflect variations in typing speed, accuracy, and style among users.
\end{itemize}

Both the outer and inner loops use a Bayesian optimization (BO) framework to guide the search for optimal parameters. We chose BO for this problem because it efficiently handles medium-dimensional and expensive-to-compute objective functions~\cite{gel_bayesian_2018}. The optimization process returns as its output the optimal general typing model and a set of human parameters, resulting in a robust typing model that performs well in various scenarios. Details of the parameters involved in the optimization can be found in the supplemental material.

\subsection{\rv{Simulation and visualization}}


Figure~\ref{fig:errorcase} gives an example of the simulation results from the model. It illustrates how errors arise and the coordination between the eyes and fingers in handling them.
Specifically, the figure depicts three sorts of text error (an omission, substitution, and insertion error), stemming from two mechanisms (lapses and slips). The first two errors have been corrected, while the third has been left uncorrected.
Such material attests that our model generates not only errors in text but also  moment-to-moment behavior in typing and fixing errors.

\begin{figure}[!t]
\centering
  \includegraphics[width=0.48\textwidth]{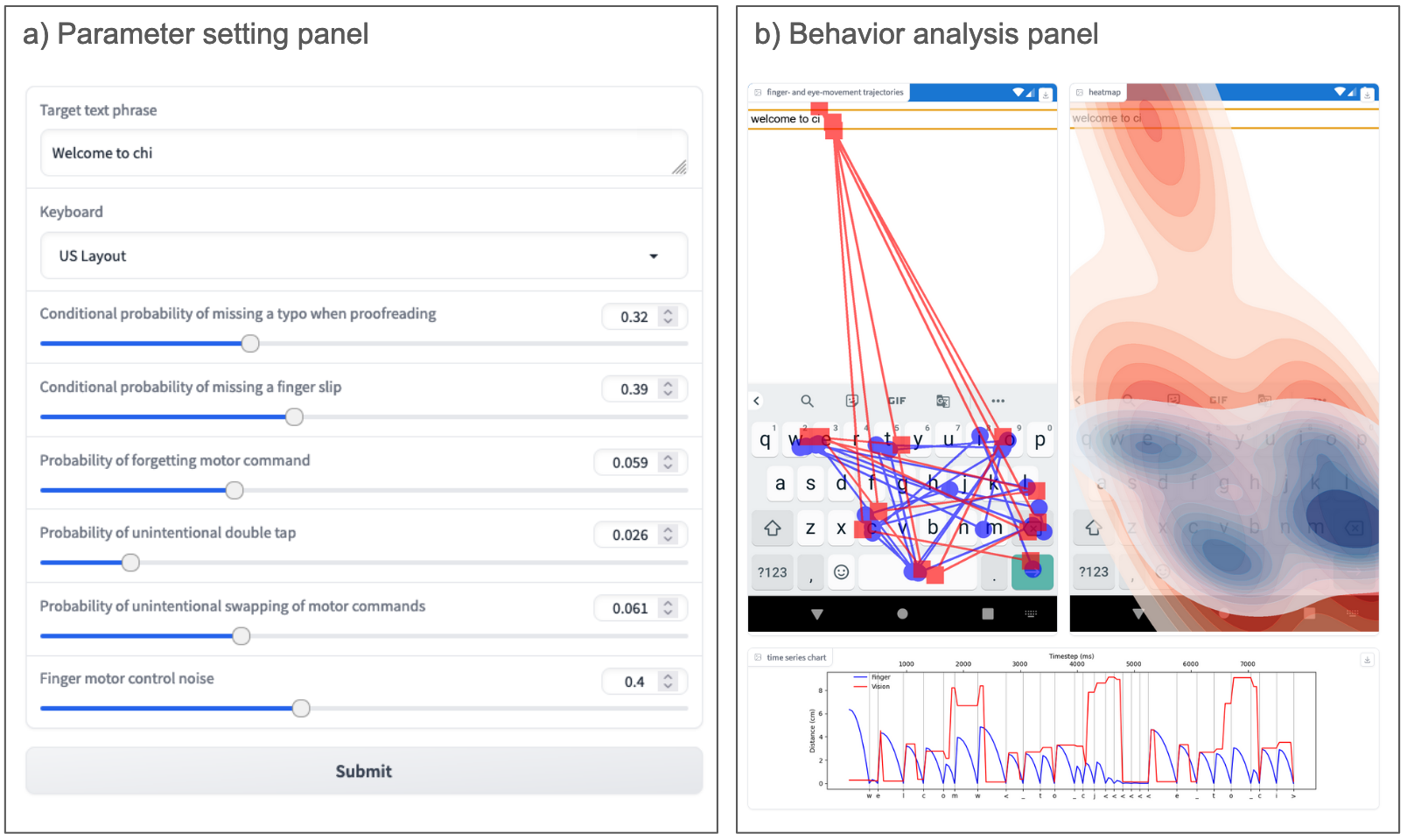}
  \caption{Visualization tool for exploring simulations. a) Via the settings panel, users can choose a target phrase for typing with the specified keyboard layout and adjust error parameters. b) The behavior analysis panel displays simulated gaze and finger movement to demonstrate the human error-linked behavior. To simulate different scenarios, the user can adjust parameters that affect the error-generating mechanisms in the model.}
  \label{fig:UI}
\end{figure}

To help practitioners and researchers simulate behaviors, we developed a visualization tool shown in Figure ~\ref{fig:UI}. 
The interface comprises a parameter setting panel (on the left) and a behavior analysis one (on the right). From the parameter setting panel, users can input a target text phrase for typing, choose a keyboard layout, and set error parameters. Upon clicking of the ``Submit'' button, the model loads the specified parameters and simulates typing behaviors, consistent with the inputs. The typing behavior generated is represented through three types of visualization:  a) A trajectory view displays the spatial movements of both gaze and finger. b) A heatmap view shows the spatial distributions of the regions traversed by the finger (in blue) and gaze (in red). c) A time series view presents the key-by-key distances from the positions of gaze and finger to the next key to tap over time, indicating the temporal relationship between the finger and the gaze. This visualization-based exploration tool allows users to fine-tune the model manually, thereby simulating human error behaviors, ones that closely match specific user performance.

\section{The \textsc{TypingError} benchmark}
\label{Sec:benchmark}

To evaluate ~\name properly, we created a benchmark incorporating datasets that capture several distinct aspects of errors in mobile typing. 
The \benchmark benchmark exhibits some overlap with the  openly available \textsc{MobileTyping} benchmark~\cite{shi2024crtypist}, but the focus here is specifically on errors. 
To that end, new datasets and metrics have been included. 
We have divided the benchmark into three ``levels'', in accordance with the constraints that study conditions may impose on errors:

\begin{itemize}
    \item \textbf{Level 0: Typing errors when errors cannot be corrected}. In this condition, , typing errors cannot be corrected. Users are asked to type as quickly and accurately as possible without making any corrections. This allows researchers to observe the full range of errors that people make.
    \item \textbf{Level 1: Typing errors when errors can be manually corrected}. In this condition, manual error corrections are allowed, with users being asked to type quickly and accurately, correcting errors upon noticing them. Backspacing is the only way of doing so.
    \item \textbf{Level 2: Typing errors when autocorrection is available}. In this condition, autocorrection of mistyped text is available, and manual error corrections are also allowed. Users can decide to correct errors themselves or rely on autocorrection.
\end{itemize}

The benchmarking presentations are arranged by level accordingly, as Table~\ref{tab:benchmark} illustrates, with corresponding datasets (see Subsec.~\ref{sec:datasets}), diverse user groups (see Subsec.~\ref{sec:user-group}), and error-related metrics (see Subsec.~\ref{sec:metrics}).

\subsection{\rv{Datasets}}
\label{sec:datasets}

\rv{
We collected human typing data from four sources~\cite{nicolau2012elderly, wang2021facilitating, palin2019people, jiang2020we}.
\begin{itemize}
    \item \textit{Parkinson's-affected text entry}~\cite{wang2021facilitating}.
    \rv{
    One dataset is centered on the text entry performance of experiment participants with Parkinson’s disease. The data collection process employed two blocks of text entry tasks, each featuring 25 phrases randomly selected from the phrase sets chosen for evaluating text entry techniques~\cite{mackenzie2003phrase}. Participants were instructed to type quickly and accurately without correcting any errors, thus affording insight into the challenges faced by individuals with motor impairments during text entry. 
    }
    \item \textit{Elderly persons' text entry}~\cite{nicolau2012elderly}.  
    \rv{
    The second dataset aids in exploring text entry performance by elderly persons and how it varies with the type of device used. To help the participants become familiar with touchscreen devices, the researchers asked them to complete tasks that involved entering single letters and copying sentences. Later in the data collection process, they asked participants to perform transcription typing tasks without correcting any errors.
    }
    \item \textit{``How We Type''}~\cite{jiang2020we}.
    \rv{
    Composed of data collected from 30 native Finnish-speakers in a controlled laboratory setting, the third dataset focuses on metrics of typing behavior at detail level. Participants were asked to type quickly and accurately such that no errors remained in the sentence submitted. The project collected eye movement data (by using SMI eye-tracking glasses) and finger motion data (through an OptiTrack Prime 13 motion-capture system).
    }
    \item \textit{``Typing37K''}~\cite{palin2019people}.
    \rv{
    The large-scale online dataset Typing37K captures transcription typing behavior from 37,000 volunteers using a Web-based platform. Participants transcribed 15 sequential sentences. Demographic data (such as age, gender, and language proficiency), typing habits, and the keyboard used were recorded also.
    }
\end{itemize}
}

\definecolor{bad}{rgb}{0.99608,0.87843,0.82353}
\definecolor{good}{rgb}{0.89804, 0.96078, 0.87843}
\definecolor{best}{rgb}{0.63137, 0.85098, 0.60784}
{\sffamily 
\begin{table*}[htbp]
\normalsize
\centering
\caption{A benchmark for evaluating model's ability to reproduce human errors in touchscreen typing in three levels of error correction. \rv{The results closest to human performance are shown in dark green. All results within 1 standard deviation from human are in light green.}  (\faEdit \  manual error correction is allowed; \faEye \  gaze data is included;) } 
\begin{tabular}{l|l|l|rrrrrr}
\hline 
\hline
\multirow{2}*{\makecell[l]{Level of error correction}} & \multirow{2}*{\makecell[l]{User Group}} &
\multirow{2}*{\makecell[l]{Metric}} & 
\multicolumn{2}{c}{Human} & \multicolumn{2}{c}{CRTypist~\cite{shi2024crtypist}} & \multicolumn{2}{c}{\name} \\
~ & ~ & ~ & $M$ & $S D$ & $M$ & $S D$ & $M$ & $S D$ \\
\hline 
\multirow{15}*{\makecell[l]{Level 0\\ \\Typing errors when errors \\cannot be corrected }} & \multirow{5}*{\makecell[l]{Young adults~\cite{wang2021facilitating} \\ \faUser \ $\times$ 8}} &WPM & 29.4 & 8.9 & - & -  & {\cellcolor{good}} 29.6 & {\cellcolor{good}} 4.2 \\
~ & ~ &Insertion errors (\%) & 0.25 & 0.08 & - & - & {\cellcolor{good}} 0.36 & {\cellcolor{good}} 1.1\\
~ & ~ &Omission errors (\%) & 0.17 & 0.08 & - & - & {\cellcolor{good}} 0.16 & {\cellcolor{good}} 0.83\\
~ & ~ &Substitution errors (\%) & 3.47 & 1.05 & - & -  &  {\cellcolor{good}} 2.50 & {\cellcolor{good}} 2.98\\
~ & ~ &Transposition errors (\%) & 0.07 & 0.04 & - & - & {\cellcolor{good}} 0.10 & {\cellcolor{good}} 0.51\\
\cline{2-9}
~ & \multirow{5}*{\makecell[l]{Parkinson’s users~\cite{wang2021facilitating} \\ \faUser \ $\times$ 8}} &WPM & 19.8 & 6.9 & - & -  & {\cellcolor{good}} 16.16 & {\cellcolor{good}} 2.05 \\
~ & ~ &Insertion errors (\%) & 6.63 & 1.24 & - & - & 2.75 & 3.4\\
~ & ~ &Omission errors (\%) & 1.13 & 0.33 & - & - & 2.24 & 3.02\\
~ & ~ &Substitution errors (\%) & 12.38 & 5.12 & - & -  & {\cellcolor{good}} 12.9 & {\cellcolor{good}} 7.74\\
~ & ~ &Transposition errors (\%) & 0.69 & 2.14 & - & - & {\cellcolor{good}} 0.67 & {\cellcolor{good}} 1.78\\
\cline{2-9}
~ & \multirow{5}*{\makecell[l]{Elderly users~\cite{nicolau2012elderly} \\ \faUser \ $\times$ 15}} &WPM & 4.70 & 3.10 & - & -  & {\cellcolor{good}} 5.60 & {\cellcolor{good}} 0.98\\
~ & ~ &Insertion errors (\%) & 4.60 & - & - & - & 0.48 & 1.8\\
~ & ~ &Omission errors (\%) & 10.80 & - & - & - & 10.12 & 11.96\\
~ & ~ &Substitution errors (\%) & 5.80 & - & - & -  & 4.82 & 5.03\\
~ & ~ &Transposition errors (\%) & 0.00 & - & - & - & 0.22 & 1.20\\
\hline 
\multirow{26}*{\makecell[l]{Level 1\\ \\Typing errors when errors\\ can be manually corrected}} & \multirow{14}*{\makecell[l]{Finnish typists~\cite{jiang2020we} \\ \faUser \ $\times$ 30 \\ \faEdit \ \faEye}} & WPM & 27.2 & 3.6 & {\cellcolor{good}}  30.2 & {\cellcolor{good}} 5.2 & {\cellcolor{best}} 26.3 & {\cellcolor{best}} 5.3 \\
~ & ~ & Uncorrected error (\%) & 0.56 & 0.71 & {\cellcolor{good}} 1.00 & {\cellcolor{good}} 2.86 & {\cellcolor{best}} 0.5 & {\cellcolor{best}} 1.57\\
~ & ~ & Corrected error (\%) & 9.38 & 5.75 & 15.40 & 11.33 & {\cellcolor{best}} 11.05& {\cellcolor{best}} 7.66\\
~ & ~ & KSPC & 1.26 & 0.37 & {\cellcolor{good}}  1.47 & {\cellcolor{good}}  0.35 & {\cellcolor{best}} 1.38 & {\cellcolor{best}} 0.29\\
~ & ~ & Backspaces & 2.61 & 1.81 & {\cellcolor{good}} 4.27 & {\cellcolor{good}}  3.45 & {\cellcolor{best}} 3.43 & {\cellcolor{best}} 3.14 \\
~ & ~ & Immediate corrections & 0.40 & 0.00 & 1.73 & 1.21 & {\cellcolor{best}} 1.47 & {\cellcolor{best}} 1.31 \\
~ & ~ & Delayed corrections & 0.63 & 0.10 & {\cellcolor{best}} 0.60 & {\cellcolor{best}} 0.66 & {\cellcolor{good}}  0.72 & {\cellcolor{good}}  0.96\\
~ & ~ &Insertion errors (\%) & 0.03 & 0.39 & - & - & {\cellcolor{good}} 0.6 & {\cellcolor{good}} 1.89\\
~ & ~ &Omission errors (\%) & 0.07 & 0.62 &  {\cellcolor{good}} 0.28 & {\cellcolor{good}}  1.5 & {\cellcolor{best}} 0.0 & {\cellcolor{best}} 0.0\\
~ & ~ &Substitution errors (\%) & 0.11 & 0.83 & {\cellcolor{good}}  0.85 & {\cellcolor{good}}  2.94 & {\cellcolor{best}} 0.0 & {\cellcolor{best}} 0.0\\
~ & ~ &Transposition errors (\%) & 0.00 & 0.18 & - & - & {\cellcolor{good}} 0.0 & {\cellcolor{good}} 0.0\\
\cline{2-9}
~ & \multirow{12}*{\makecell[l]{Gboard typists~\cite{palin2019people}\\\faUser \ $\times$ 5,140 \\ \faEdit}} & WPM & 35.7 & 13.8 & {\cellcolor{good}}  30.57 & {\cellcolor{good}} 5.75 & {\cellcolor{best}} 38.02 & {\cellcolor{best}} 10.32 \\
~ & ~ & Uncorrected error (\%) & 3.44 & 3.79 & {\cellcolor{good}}  0.7 & {\cellcolor{good}} 6.59 & {\cellcolor{best}} 5.56 & {\cellcolor{best}} 4.4\\
~ & ~ & Corrected error (\%) & 4.95 & 6.71 & 12.77 & 9.92 & {\cellcolor{best}} 5.69 & {\cellcolor{best}} 6.76\\
~ & ~ & KSPC & 1.13 & 0.19 & 1.38 & 0.29 & {\cellcolor{best}} 1.34 & {\cellcolor{best}} 0.43\\
~ & ~ & Backspaces & 2.42 & 3.61 & {\cellcolor{good}} 3.87 & {\cellcolor{good}} 3.42 & {\cellcolor{best}} 3.6 & {\cellcolor{best}} 4.62 \\
~ & ~ & Immediate corrections & 0.64 & 0.93 & {\cellcolor{good}} 1.37 & {\cellcolor{good}} 1.11 & {\cellcolor{best}} 0.67 & {\cellcolor{best}} 0.81 \\
~ & ~ & Delayed corrections & 0.53 &  0.89 & {\cellcolor{best}} 0.83 & {\cellcolor{best}} 1.0 & {\cellcolor{good}} 0.96 & {\cellcolor{good}} 1.37 \\
~ & ~ &Insertion errors (\%) & 0.97 & 1.99 & - & - & {\cellcolor{good}} 0.02 & {\cellcolor{good}} 0.31\\
~ & ~ &Omission errors (\%) & 0.94 & 1.97 & {\cellcolor{good}} 0.0 & {\cellcolor{good}} 0.0 & {\cellcolor{best}} 0.06 & {\cellcolor{best}} 0.57\\
~ & ~ &Substitution errors (\%) & 1.78 & 2.72 & {\cellcolor{best}} 0.79 & {\cellcolor{best}} 1.65 & 6.38 & 5.1\\
~ & ~ &Transposition errors (\%) & 0.08 & 0.49 & - & - & {\cellcolor{good}} 0.01 & {\cellcolor{good}} 0.28\\
\hline 
\multirow{12}*{\makecell[l]{Level 2\\ \\Typing errors when\\  autocorrection is available}} & \multirow{12}*{\makecell[l]{Auto-correction\\users~\cite{palin2019people}\\\faUser \ $\times$ 148 \\ \faEdit}} & WPM & 32.2 & 12.0 & {\cellcolor{best}} 36.89 & {\cellcolor{best}} 7.74 & {\cellcolor{good}} 38.38 & {\cellcolor{good}} 11.9 \\
~ & ~ & Uncorrected error (\%) & 3.39 & 4.15 & {\cellcolor{good}} 1.73 & {\cellcolor{good}} 5.08 & {\cellcolor{best}} 2.32 & {\cellcolor{best}} 5.45\\
~ & ~ & Corrected error (\%) & 3.48 & 6.08 & {\cellcolor{best}} 3.59 & {\cellcolor{best}} 6.62 &  {\cellcolor{good}} 5.33 & {\cellcolor{good}} 7.39\\
~ & ~ & KSPC & 1.09 & 0.19 & {\cellcolor{best}} 1.34 & {\cellcolor{best}} 1.24 & {\cellcolor{good}} 1.37 & {\cellcolor{good}} 0.47\\
~ & ~ & Backspaces & 1.80 & 3.09 & {\cellcolor{best}} 1.59 & {\cellcolor{best}} 4.9 & {\cellcolor{good}} 3.35 & {\cellcolor{good}} 4.62\\
~ & ~ & Immediate corrections & 2.10 & 3.22 & {\cellcolor{good}} 2.97 & {\cellcolor{good}} 13.7 & {\cellcolor{best}} 1.72 & {\cellcolor{best}} 3.68 \\
~ & ~ & Delayed corrections & 0.47 & 0.93 & {\cellcolor{good}} 0.05 & {\cellcolor{good}} 0.21 & {\cellcolor{best}} 0.61 & {\cellcolor{best}} 1.11 \\
~ & ~ &Insertion errors (\%) & 1.06 & 2.17 & - & - & {\cellcolor{good}} 0.08 & {\cellcolor{good}} 0.91\\
~ & ~ &Omission errors (\%) & 1.01 & 1.69 & {\cellcolor{best}} 0.41 & {\cellcolor{best}} 3.08 & {\cellcolor{good}} 0.31 & {\cellcolor{good}} 2.67 \\
~ & ~ &Substitution errors (\%) & 1.58 & 3.57 & {\cellcolor{best}} 1.9 & {\cellcolor{best}} 5.61 & {\cellcolor{good}} 2.28 & {\cellcolor{good}} 4.45 \\
~ & ~ &Transposition errors (\%) & 0.08 & 0.50 & - & - & {\cellcolor{good}} 0.02 &{\cellcolor{good}}  0.23\\
\hline
\hline
\end{tabular}
\label{tab:benchmark}
\end{table*}
}


\subsection{User groups}
\label{sec:user-group}

\rv{
The user groups were derived from the four datasets, with the data for each group being broken down further by our three levels.
}

At \textbf{Level 0}, the data we have includes the typing activity for individuals who were using a touchscreen without making any corrections. The three sets of users were 
\begin{enumerate}
    \item A group consisting of eight young adults (5 female and 3 male, all right-handed), with an average age of 23.6 years (standard deviation (\emph{SD}) = 3.7)~\cite{wang2021facilitating} 
    \item Eight Parkinson's patients (3 female and 5 male, all right-handed), 60.5 years old on average ({SD} = 9.2, with a range of 47 to 72), from a Parkinson's foundation~\cite{wang2021facilitating}
    \item Fifteen participants (11 female and 4 male), with ages ranging from 67 to 89 and a mean age of 79 (standard deviation = 7.3)~\cite{nicolau2012elderly}
\end{enumerate}

At \textbf{Level 1}, we used data from two separate keyboard layouts: an English and a Finnish one. 
For the Finnish-layout keyboard, we used material from the How We Type dataset~\cite{jiang2020we}, from 30 native Finnish-speakers with normal or corrected vision.
For the English-layout one, we selected a subset from Typing37K~\cite{palin2019people} (5,140 typing trajectories) in which participants were using the Gboard interface and typing without any intelligent features. Since the data were collected from an online-test Web site, participants were more careless but faster than those in the laboratory study.

At \textbf{Level 2}, we further refined the human data from Typing37K by filtering out data with participants using the Gboard interface with \emph{only} autocorrection. This left us with 148 typing trajectories.

\subsection{Error-related metrics}
\label{sec:metrics}

While including general typing metrics such as the commonly used words per minute (WPM) speed measurement, obtained by calculating the number of words divided by the time taken, our benchmark places more emphasis on error-related metrics.

\begin{itemize}
    \item \textit{Uncorrected error rate}~\cite{wobbrock2007measures}: The percentage of non-corrected incorrect keystrokes over the total of incorrect and correct keystrokes.
    \item \textit{Corrected error rate}~\cite{wobbrock2007measures}: Incorrect but rectified keystrokes as a percentage of the  sum of incorrect plus correct keystrokes.
    \item \textit{Keystrokes per character}~\cite{wobbrock2007measures}: The number of keystrokes divided by the number of characters produced (a larger number indicates more corrections).
    \item \textit{Backspaces}~\cite{palin2019people}: The number of \texttt{Backspace} presses for error correction during the typing of the text.
    \item \textit{Immediate error corrections}~\cite{arif2016evaluation}: This refers to the frequency of error correction in which the user immediately identifies and corrects an error with a subsequent Backspace press.
    \item \textit{Delayed error corrections}~\cite{arif2016evaluation}: This denotes the frequency of error correction wherein the user tries to correct previously missed errors in the middle of the text.
    \item \textit{Insertion error rate}~\cite{wang2021facilitating}: The rate of redundant touches that do not correspond to any of the target characters.
    \item \textit{Omission error rate}~\cite{wang2021facilitating}:  The rate of characters that do not correspond to any of the input touch points.
    \item \textit{Substitution error rate}~\cite{wang2021facilitating}: The rate of touches intended for certain characters, but landed on different keys.
    \item \textit{Transposition error rate}~\cite{wang2021facilitating}: The rate of touches resulting in characters being swapped.
\end{itemize}
\section{Results}

This section presents the results of our evaluation via the benchmark, presented in Table~\ref{tab:benchmark}.
In general, our model can synthesize diverse errors based on the cognitive mechanisms we incorporated. The model can closely reproduce human-like behavior at each level of error correction in the benchmark. In the following section, we will analyze the results in detail for each level.

\subsection{Level 0: Typing errors when errors cannot be corrected}

To train a model set-up without error correction, we disabled the \texttt{Backspace} key in the internal environment, so as to simulate typing without the ability to make corrections. The goal of the model remained correct typing of the phrase. However, the model could only control gaze movement to guide finger placement for correct typing and check the text field to determine the next input.

We evaluated the model by considering three user groups. After adjusting cognitive parameters for these groups, we ran simulations with the same number of independent episodes for each group, then compared the results. The model's generation of errors here, shown under Level 0 in Table \ref{tab:benchmark}, can be characterized thus:

\begin{itemize}
    \item[1)] \textit{Young adults}: The model can accurately reproduce the typing errors made by young adults. In typing at a speed of approx. 29~WPM, all types of errors fall within one standard deviation of the data for young adults. The error rates' prevalence order matches that of the humans: substitution errors, insertion errors, omission errors, transposition errors. 
    \item[2)] \textit{Users with Parkinson's}: The model also generates results similar to those of the Parkinson's patients. When typing at a (slow) speed close to that of this group, it yields nearly identical substitution and transposition errors. Also, it accurately reproduces the order of error rates within this group too, with substitution errors being the most common, followed by insertion, omission, and transposition errors. However, the model produces fewer insertion errors than seen in the data from actual users with Parkinson's. In this case, the probability of unintentional double tapping is much higher than the model expects.
    \item[3)] \textit{Elderly users}: Elderly users are the group with the slowest typing. Our model still is able to account for that typing speed within one standard deviation. As for individual error classes, the reference data do not include the standard deviation for each, but if we assume the \emph{SD} to be 1\%, three of the error rates from the model fall within that range, while insertion errors constitute an exception. The most common error type among elderly users is omission errors due to forgetfulness in cognition, and the model can closely match that phenomenon. It also successfully replicates the order of error rates within this group: omission errors dominate, followed by substitution errors, then insertion errors, and finally transposition errors.
\end{itemize}

\subsection{Level 1: Typing errors when errors can be manually corrected}

\begin{figure*}[!t]
\centering
  \includegraphics[width=\textwidth]{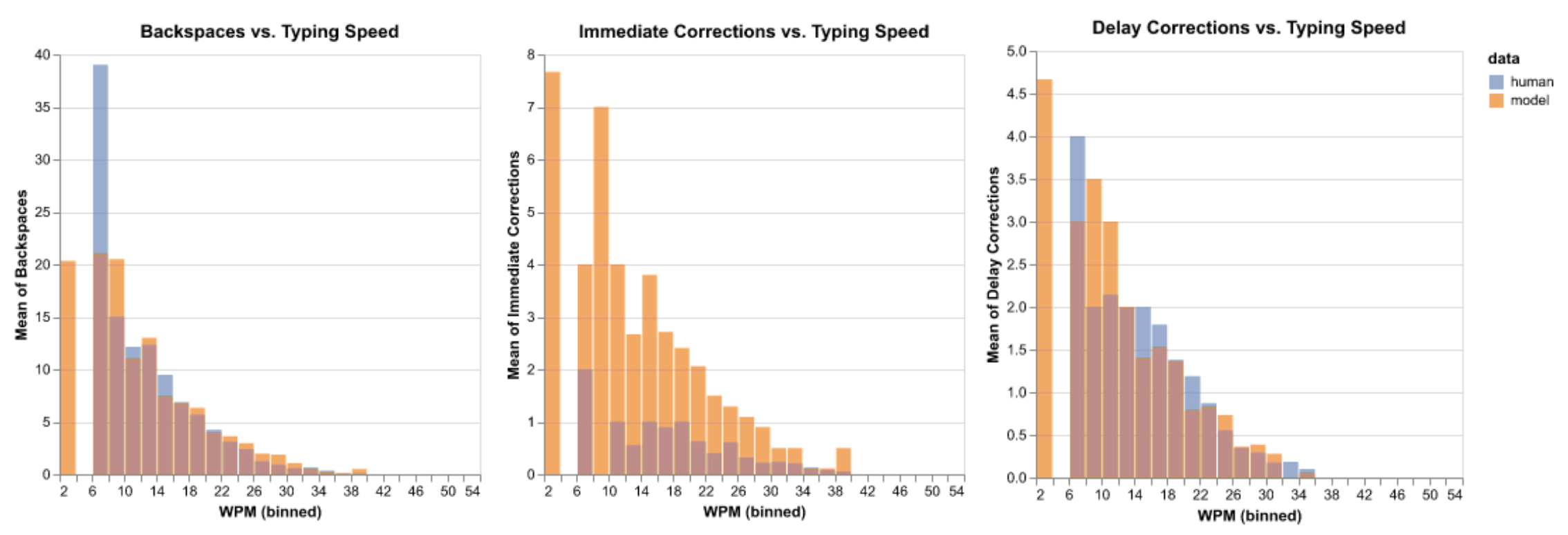}
  \caption{Typing speed vs. error corrections. The figure shows the speed--accuracy tradeoff in both human data and the predictions.
}
  \label{fig:corrections_wpm}
  \vspace{-3mm}
\end{figure*}

In the conditions at this level, users were encouraged to improve the accuracy of the typed text. Therefore, the model's internal environment included the goal for the \texttt{Backspace} key within the action space. The model was trained to type phrases quickly and accurately, and optimization of all human parameters for the model for the target user group was handled by minimizing the differences in typing speed, error rates, and the amount of backspacing.

In the Finnish typing dataset, we sampled 30 independent runs similar to the collected human data. Our model with optimized parameters demonstrated human-like error-handling strategies. Compared to the state-of-the-art baseline approach~\cite{shi2024crtypist}, our model showed similar performance for uncorrected-error rate but proved much closer to humans in its corrected-error rate. Relative to the baseline, \name also uses the \texttt{Backspace} key in a more human manner here, resulting in a similar number of keystrokes per character. As Table~\ref{tab:benchmark} indicates, the baseline model performs better only for the number of delayed corrections, and even for these our model stays within a standard deviation of humans.
\rv{
The model types carefully, in line with human behavior, so produces few errors in the final sentences submitted. In our experiment, the model corrected all omission, substitution, and transposition errors, reaching accuracy levels close to human performance: the average was 0.07\% for omission errors, the rate was 0.11\% for substitution errors, and no transposition errors were observed.
}

We found that the relationship between \name's error corrections and typing speed was consistent with the distinct error correction patterns of users. Since the human parameters are normalized from 0 to 1, we sampled them from a Gaussian distribution, with the mean representing the optimal parameter and a standard deviation of 0.1. We carried out 300 independent simulations using Finnish sentences from the dataset~\cite{jiang2020we} to explore the connection between error correction and typing speed. The results (shown in Figure~\ref{fig:corrections_wpm}) showcase how the model can capture the distribution of errors and replicate the speed--accuracy preference observed in human users. The only difference is that the model made twice as many immediate corrections as the humans did; i.e., it shows a tendency to correct errors immediately.

In our work with the Gboard data, we ran 5,140 independent simulations with the model, matching the number of trials in the human data. Our model can replicate careless behavior, exhibiting a relatively high uncorrected error rate, yet also yields a corrected error rate consistent with human users'. Additionally, it exhibits error correction behavior that is similar to humans' (lying within one standard deviation of the human data) by all metrics, except for the substitution-error rate. The model tends to leave more substitution errors in the text submitted. In a parallel to the test with the Finnish typing dataset, our model showed a slightly elevated number of delayed corrections, but it was still within one standard deviation of the human data.

\rv{
To evaluate how closely the synthesized data from \name and CRTypist align with the human data's distribution, we conducted statistical analysis using the Bayes factor from a \textit{t}-test function~\cite{rouder2009bayesian} utilizing the \texttt{Pingouin} library~\footnote{\textit{\url{https://pingouin-stats.org/}}}. For each model (\name and CRTypist), we simulated 30 data points, thus matching the number of human data. 
We defined the null hypothesis ($H_0$) as no difference between the simulated data and human data, while the alternative hypothesis ($H_1$) asserts that a difference does exist.
The test results suggest that \name aligns more closely with human data than does CRTypist, across most metrics, for both the Finnish typing dataset (\name: 6/7 show support for $H_0$; CRTypist: 2/7 support $H_0$) and the Gboard dataset (\name: 2/7 support $H_0$; CRTypist: 0/7 support $H_0$).
From our tests with the Finnish typing dataset, the results for \name indicate support for $H_0$ by nearly all metrics, including metrics: WPM ($BF_{10} = 0.576$), uncorrected error rate ($BF_{10} = 0.382$), corrected error rate ($BF_{10} = 0.302$), and KSPC ($BF_{10} = 0.593$), delayed corrections ($BF_{10} = 0.529$). The only exception is immediate corrections ($BF_{10} = 81.015$). In contrast, $H_0$ with CRTypist receives support from only two metrics: uncorrected error rate ($BF_{10} = 0.262$) and delayed corrections ($BF_{10} = 0.529$).
With the Gboard dataset, $H_0$ is likewise strongly supported for \name. Alignment is excellent for WPM, corrected-error rate, backspacing, immediate corrections, and delayed corrections. In contrast, the CRTypist evidence supports $H_1$ across all metrics. Detailed analysis results can be found in the supplemental material.
}

\subsection{Level 2: Typing errors when autocorrection is available}

For the final level, at which autocorrection is used when the user types text, we improved the external environment by incorporating a rule-based feature that automatically corrects a word if its edit distance from the target word is within two characters. This autocorrection is triggered once the space bar is pressed.

We executed 148 independent runs of the model, which exhibited a slight increase in typing speed when autocorrection is not enabled. This enhancement not only reduced the uncorrected error rate by half but also slightly decreased the corrected error rate. This suggests fewer instances needing manual correction -- a conclusion supported by the reduced backspacing -- while higher accuracy is maintained. Notably, the model favored immediate corrections over delayed ones, relying on the autocorrection feature to rectify earlier mistakes.

\rv{
Research indicates that improvements in typing speed are linked to the accuracy of any autocorrection features~\cite{roy2021typing}. However, contrary to expectations and findings from previous studies~\cite{banovic2019limits}, our source human data demonstrate a decrease in typing speed when autocorrection is enabled. 
}
This discrepancy might be attributable to complexities encountered in real-world typing conditions.
For instance, the simulation of autocorrection might overlook some errors created/exacerbated by autocorrection itself. For instance, in an error type known as ``space key confusion'', users accidentally hit the space bar instead of producing the intended non-space character, thus triggering unintended autocorrection and the insertion of incorrect words.





\section{Discussion}

\rv{
\name is the first computational model to accurately simulate a wide range of human errors in a complex, real human--computer interaction task.
Specifically, it simulates omission, transposition, commission, and substitution errors in typing.
The model achieves a high level of similarity with human data across multiple conditions and groups, both as judged via aggregate metrics, such as WPM, and when handling trajectory-level predictions.
}

\rv{
What do the results mean for practitioners and for broader understanding of human errors, though, and what work remains to be done?
To tackle these key questions, we discuss the implications and limitations of the results next.
}

\subsection{\rv{Implications}}

\rv{
We see three exciting avenues in applying ~\name: evaluation, user research, and generation of synthetic data.
}

\rv{
Firstly,  
\name makes it possible to evaluate keyboard designs before undertaking an empirical study of them. Compared to CRTypist, \name generates more realistic error patterns and error-handling behavior; hence, it proves more effective for evaluating the fault tolerance of a given design.
It is valuable in covering more errors too, because seemingly innocuous aspects of a design can have surprising effects downstream, on users.
Errors take lots of time to spot and correct during typing; hence, minimizing their occurrence is a major aim in the design of any text entry system.
}

\rv{
Secondly, 
~\name enables practitioners to study individual-level differences in typing. 
The results presented under Level 0 in Table~\ref{tab:benchmark} attest to \name's ability to reproduce diverse error patterns from elderly individuals and users with Parkinson's disease~\cite{nicolau2012elderly, wang2021facilitating}. This is thanks to the explainable modular architecture,
which can support varying the free parameters for vision, motor, and memory that constrain the cognitive capacities of the model. 
We conclude, then, that the architecture design underpinning \name displays potential to generate error behaviors consistent not only with ``average'' users but also with specific target groups with unique characteristics.
}

\rv{
Thirdly,
intelligent text entry (ITE) techniques often rely on supervised learning.
We believe that, on account of the realistic nature of its predictions, ~\name affords new methods of data augmentation,
wherein synthetically produced data serve to complement a dataset, particularly in conditions where empirical data may be hard to collect.
}

\rv{
Looking beyond practical applications, we find the model to hold promise for opening the door to a new way of theorizing about errors in human--computer interaction.
The results of our work stem from a single key assumption behind our model: that users can strategically allocate resources to monitor and correct errors.
This complements the prevailing understanding of human errors, which has focused on the mechanisms that generate errors but not those that fix them. 
The underlying principle is aligned with the nascent theory of resource rationality \cite{lieder2020resource}, according to which people adaptively control the way they use their cognition.
From an RL perspective, they learn policies on their cognitive machinery -- and not just for their overt behavior.
Our computational implementation lends credence to this idea, as do the results obtained. 
}

\subsection{\rv{Limitations and Future Work}}

\rv{
Much is yet to be done to extend ~\name to support the many types of intelligent features developed for keyboards today.
At present, \name does not completely capture real-world behaviors when autocorrection is involved. We noticed that some errors stem from conflicting correction mechanisms. In this case, the autocorrecting operation may intervene at the very moment the user is trying to correct a mistake. Such simultaneous execution can lead to situations wherein a ``bad correction'' is made, due not to human error but, rather, a misalignment between the user’s act and the automated system’s action. Future efforts must consider dynamic interactions such as these between user inputs and intelligent feedback.
}

\rv{
We readily acknowledge that real-world behavior with ITE techniques is more complex than what our model currently encompasses at ``Level 2.'' 
\name should be extended to handle commonly used techniques for interactively correcting errors, such as selecting text in a modal manner (e.g., with a ``caret''), 
gesture-based text entry~\cite{zhai2003shorthand}, and more advanced techniques \cite{zhang2019type}.
One of the most popular features employed in modern typing is word prediction, which has become integral to the typing process across both mobile and desktop environments.
Word prediction systems allow users to select suggested words, hence bypassing both traditional typing and error correction mechanisms. However, \name does not yet cover how predictive features of such a nature influence typing and its correction. Bridging this gap could be the fruit of future work that implements the latest features in the training environment.
}

\section{Conclusion}

\rv{
To sum up, this paper contributes a computational model for simulating errors in touchscreen typing. 
%
By generating realistic distributions for four typographical error types,
covering a wide range of individual differences, 
and adapting to the complicated case of autocorrection during typing, 
\name demonstrated state-of-the-art performance in challenging benchmarking for reproduction of human errors in typing.
The model did all this without compromising its strong performance by other important metrics for typing.
We conclude that these results point to great potential in the class of models whereby the prediction of user behavior is rooted in maximizing expected utility under cognitive bounds.
This approach marks a notable divergence from the data-driven approaches so popular today:
in explicitly modeling the \emph{causes} of errors, instead of just ``parroting'' statistically plausible typographical errors in text, 
the model takes a \emph{glass-box} rather than a black-box approach. Every typographical error can be traced to the underlying cognitive events that produced it. 
However, the current version of the model does not fully account for real-world behaviors involving advanced features. Future work should aim to enhance the model to better handle complex dynamics such as gesture-based text input and word prediction.
}


\begin{acks}
This work was supported by 
the Research Council of Finland project Subjective Functions (grant 357578), 
Finnish Center for Artificial Intelligence (grants 328400, 345604, 341763),
European Research Council Advanced Grant (no. 101141916),
the Department of Information and Communications Engineering at Aalto University, 
and Google donation (DeepTypist).
The calculations were performed via computer resources provided by the Aalto University School of Science project Science-IT. The authors also acknowledge Finland's CSC – IT Center for Science Ltd. for providing generous computational resources.
\end{acks}

\bibliographystyle{ACM-Reference-Format}
\bibliography{reference}

\appendix











\end{document}